\documentclass[fleqn,usenatbib]{mnras}

\usepackage[T1]{fontenc}

\DeclareRobustCommand{\VAN}[3]{#2}
\let\VANthebibliography\thebibliography
\def\thebibliography{\DeclareRobustCommand{\VAN}[3]{##3}\VANthebibliography}

\usepackage{graphicx}	
\usepackage{amsmath}	
\usepackage{amssymb}	
\usepackage{subcaption}
\usepackage[dvipsnames]{xcolor}
\usepackage{xspace}
\setcitestyle{notesep={ }}
\usepackage{comment}

\newcommand{\mufasa}{{\sc Mufasa}\xspace}
\newcommand{\simba}{{\sc Simba}\xspace}

\newcommand{\lya}{Ly$\alpha$\xspace}
\newcommand{\HI}{\ion{H}{i}\xspace}
\newcommand{\MgII}{\ion{Mg}{ii}\xspace}
\newcommand{\CII}{\ion{C}{ii}\xspace}
\newcommand{\SiIII}{\ion{Si}{iii}\xspace}
\newcommand{\CIV}{\ion{C}{iv}\xspace}
\newcommand{\OVI}{\ion{O}{vi}\xspace}

\newcommand{\hkpc}{h^{-1}{\rm kpc}}

\newcommand{\hmpc}{h^{-1}{\rm Mpc}}

\newcommand{\kms}{\;{\rm km}\,{\rm s}^{-1}}

\newcommand{\msolar}{\;{\rm M}_{\odot}}

\newcommand{\gizmo}{{\sc Gizmo}\xspace}
\newcommand{\caesar}{{\sc Caesar}\xspace}

\newcommand{\mstar}{\mbox{$M_\star$}\xspace}

\newcommand{\pygad}{{\sc Pygad}\xspace}

\newcommand{\cloudy}{{\sc Cloudy}\xspace}

\def\response{\textcolor{black}}

\title[Machine learning and the CGM]{Mapping Circumgalactic Medium Observations to Theory Using Machine Learning}

\author[S. Appleby et al.]{
Sarah Appleby$^{1}$\thanks{E-mail: sarahappleby20@gmail.com},
Romeel Dav\'e$^{1,2,3}$,
Daniele Sorini$^{1, 4, 5}$,
Christopher C. Lovell$^{6}$,
Kevin Lo$^{1}$
\\
\\$^{1}$ SUPA\thanks{Scottish Universities Physics Alliance}, Institute for Astronomy, University of Edinburgh, Royal Observatory, Edinburgh EH9 3HJ, UK
\\$^2$ University of the Western Cape, Bellville, Cape Town 7535, South Africa
\\$^4$ D\'epartement de Physique Th\'eorique, Universit\'e de Gen\`eve, 24 quai Ernest Ansermet, 1211 Gen\`eve 4, Switzerland
\\$^5$ Institute for Computational Cosmology, Department of Physics, Durham University, South Road, Durham, DH1 3LE, UK
\\$^6$ Institute of Cosmology and Gravitation, University of Portsmouth, Burnaby Road, Portsmouth, PO1 3FX, UK}

\date{Accepted XXX. Received YYY; in original form ZZZ}

\pubyear{2023}

\begin{document}
\label{firstpage}
\pagerange{\pageref{firstpage}--\pageref{lastpage}}
\maketitle

\begin{abstract}
We present a random forest framework for predicting circumgalactic medium (CGM) physical conditions from quasar absorption line observables, trained on a sample of Voigt profile-fit synthetic absorbers from the \simba cosmological simulation. Traditionally, extracting physical conditions from CGM absorber observations involves simplifying assumptions such as uniform single-phase clouds, but by using a cosmological simulation we bypass such assumptions to better capture the complex relationship between CGM observables and underlying gas conditions.
We train random forest models on synthetic spectra for \HI and selected metal lines around galaxies across a range of star formation rates, stellar masses, and impact parameters, to predict absorber overdensities, temperatures, and metallicities.
The models reproduce the true values from \simba well, with \response{normalised transverse standard deviations of $0.50-0.54$ dex in overdensity, $0.32-0.54$ dex in temperature, and $0.49-0.53$ dex in metallicity predicted from metal lines (not \HI), across all ions.}
Examining the feature importance, the random forest indicates that the overdensity is most informed by the absorber column density, the temperature is driven by the line width, and the metallicity is most sensitive to the specific star formation rate.
Alternatively examining feature importance by removing one observable at a time, the overdensity and metallicity appear to be more driven by the impact parameter.
We introduce a normalising flow approach in order to ensure the scatter in the true physical conditions is accurately spanned by the network.
The trained models are available online.
\end{abstract}




\section{Introduction}

Over recent years, there has been much effort to characterise the CGM via quasar absorption line studies \citep[see reviews by][]{putman_2012, tumlinson_2017, peroux_2020}. Many of the studies probe the strong transitions that exist in the rest ultra-violet (UV) regime and which trace cool or warm gas. 
Such studies are motivated by the wish to understand the baryon cycle of gas flows in the CGM: accretion onto galaxies from the IGM and satellite galaxies; expulsion of gas via stellar winds and AGN feedback; recycling of previously ejected material back onto galaxies.

The physical conditions of the CGM are studied by retrieving kinematics, spatial distributions, metallicities, densities, and temperatures from the absorption features \citep[e.g. ][]{stocke_2013, savage_2014, werk_2014, lehner_2014, lehner_2018, lehner_2019, wotta_2016, wotta_2019, keeney_2017, prochaska_2017, qu_2022}. 
To extract physical conditions, absorption systems are commonly fitted with Voigt profiles to model each absorption component and obtain column densities, linewidths and redshift-space positions. By running ionisation models (typically using {\sc Cloudy}, \citealt{ferland_2017}) and varying the input physical parameters, a Bayesian search can be performed across parameter space for the physical conditions of each absorber component using the ensemble of absorption properties as constraints. In such models the clouds are often modelled as plane-parallel slabs of gas with an ionising flux incident on one face, making the (simplifying) assumption that each cloud is spatially isolated with single-valued properties \citep[e.g.][]{churchill_2003, tripp_2008, werk_2014, fumagalli_2016, keeney_2017, prochaska_2017}.

The analysis and interpretation of CGM observations poses many challenges owing to the complex nature of the halo environment. The shapes of absorption profiles are sensitive to the underlying phase structure and likely contain contributions from different phases, for example due to the motion of gas within the halo and the clumpy gas structure.
Even within individual absorber systems the metallicity of the absorbing gas can vary and multiple gas phases may be present \citep{lehner_2019, zahedy_2019, sankar_2020, chen_2020, haislmaier_2021, sameer_2021}.
Detailed analysis of absorption systems can give relative abundances of different ions that constrain the physical conditions, but this requires high resolution spectroscopy. Studies that use this technique have moved away from the assumption of a single cloud, by modelling the high and low excitation ions separately \citep{zahedy_2019, zahedy_2021, haislmaier_2021, qu_2022}, or by modelling the absorption components as arising from multiple clouds \citep{cooper_2021, sameer_2021, nielsen_2022}. Interpreting the observational picture is further complicated due to the sensitivity of density and metallicity estimates to the shape of the UVB \citep{oppenheimer_2013, acharya_2022, gibson_2022}.
Furthermore, particular ions are not necessarily produced by the same structures and processes at different redshifts due to the evolving UVB \citep{haardt_2012, faucher-giguere_2020}. 

Galaxy formation simulations provide a valuable theoretical perspective on these problems as they offer complete particle data and physical properties for the gas that makes up the CGM, making it possible to directly interpret observations. A range of UV metal lines have been used to probe the cool and warm ionised CGM in simulations, testing specific stellar wind implementations \citep{ford_2013, ford_2014, ford_2016, hummels_2013}, the NIHAO simulation suite \citep{gutcke_2017}, EAGLE \citep{oppenheimer_2016, oppenheimer_2018b}, IllustrisTNG \citep{nelson_2020, defelippis_2021}, FIRE-2 \citep{li_2021} and \simba \citep{appleby_2021, appleby_2023}.

Such simulations can be useful for examining the impact of different line analysis methods on the retrieved CGM gas conditions \citep[e.g.][]{churchill_2015, liang_2018}.
In a recent analysis of a sample of synthetic absorption lines from a cosmological simulation, \cite{marra_2021} tested the accuracy of the single cloud ionisation modelling method of retrieving physical gas conditions. The authors find that while there is general agreement between intrinsic conditions and those derived from ionisation modelling, such methods capture the average properties of absorbing gas cells, consistent with observational tests by \cite{sameer_2021} comparing single-phase and multiphase modelling. 
\cite{marra_2022} followed up by testing the assumption of single spatially-isolated absorbing clouds in the CGM, showing that several distinct absorbing clouds may be present within a single absorption component. The distinct clouds may arise from gas of different phases that happen to be aligned kinematically. These results demonstrate that the CGM is a complex environment, with non-linear relationships between the underlying CGM conditions and the resulting absorption observables.

Machine learning (ML) algorithms have the capacity to learn complex, non-linear relationships and as such they have been widely applied to astrophysical problems \citep[see review by][]{fluke_2020}. 
In this paper, we explore a novel approach for cosmological simulations to aid in interpreting CGM absorption observations using ML models. 
We present a framework for Random Forest (RF) mapping between synthetic CGM absorption observables from the \simba\ simulation \citep{dave_2019} and the underlying absorber conditions from particle data. Such a mapping has the potential to be employed as a useful tool in retrieving physical conditions from real, multi-component absorption observations.
This approach eliminates the need for simplifying assumptions about the structure and state of the gas, i.e. whether absorption arise from single or multiple gas phases. Instead the RF mappings implicitly assume the veracity of the \simba\ galaxy formation model and our choice of UVB \citep{faucher-giguere_2020} to produce its predictions.  

The \simba\ simulations accurately reproduce a variety of observational galaxy properties. At low redshift, these include the star-forming main sequence, black hole-galaxy co-evolution, radio galaxy populations, dust properties, cold gas properties, and the baryonic Tully-Fisher relation \citep[][]{dave_2019, dave_2020, thomas_2019, thomas_2021, li_2019, lovell_2021, glowacki_2020, appleby_2020}. On larger mass scales, \simba\ reproduces X-ray scaling relations for massive halos \citep{robson_2020} and low redshift \lya\ absorption statistics of the IGM \citep{christiansen_2020}.
In previous work we have shown that \simba\ also broadly reproduces the observed absorption properties of \HI\ \citep{sorini_2020} and selected metal lines in the CGM \citep{appleby_2021}, and that such absorption arises from physically reasonable gaseous conditions \citep{appleby_2023}; therefore \simba\ is a reasonable choice of simulation with which to explore the capabilities of ML methods to learn relationships in the CGM. 

Nonetheless, there is no guarantee \simba\ yields fully accurate and representative circum-galactic media. 
Indeed, CGM zoom simulations suggest that \simba's resolution may be too poor to capture finer details of multi-phase gas, particularly for stronger absorbers ~\citep[e.g.][though see \citealt{nelson_2020}]{van_de_voort_2019,suresh_2019}.
This drawback could be explored via comparing the results of this framework applied to other simulations.  
We leave this aspect for future work, and here focus on presenting the general framework and its results when applied to the \simba\ model.

In this paper we train Random Forest (RF) machine learning networks on the low-redshift \simba\ CGM absorber sample presented in \cite{appleby_2023} to produce predictions for the underlying gas conditions in the CGM. 
This paper is organised as follows. 
In \S\ref{sec:sims} we present the \simba simulations. In \S\ref{sec:absorbers} we describe the galaxy selection, spectrum generation and fitting processes.
In \S\ref{sec:random_forest} we describe the Random Forest (RF) model and training process. In \S\ref{sec:ml_accuracy} we examine the accuracy of the RF models. In \S\ref{sec:features} we examine the feature importance of the RF models. In \S\ref{sec:phase_space} we present the RF predictions in phase space. Finally in \S\ref{sec:conclusions} we conclude and summarise. 

\section{Simulations}\label{sec:sims}

\simba \citep{dave_2019} is a suite of state-of-the-art cosmological simulations that is the successor to the \mufasa\ simulations \citep{dave_2016}, with the major additions being the inclusion of two-mode black hole growth and three-mode black hole feedback, along with an on-the-fly dust evolution model.  The main simulation, and the one employed in this work, contains $1024^3$ gas cells and the same number of dark matter particles within a ($100\hmpc$)$^3$ volume.  This yields a particle mass resolution of $1.8\times 10^7 M_\odot$ per gas cell, and $9.6\times 10^7 M_\odot$ per dark matter particle, with a spatial resolution of $\approx 1\hkpc$ in the densest regions.  Since \simba\ has been extensively described in many previous works, and since the primary goal on this work is to present and explore our machine learning framework that is not crucially dependent on which simulation it is applied to, for brevity we do not present all of \simba's input physics, but rather refer readers to \cite{dave_2019}, \cite{thomas_2019} and \cite{li_2019} for full details.

\section{Absorber sample}\label{sec:absorbers}

\begin{table*}
    \centering
    \begin{tabular}{c|c|c|c|c|c}
        \hline
        Species & n & ${\rm log }(N_{\rm min}/ {\rm cm}^{-2})$ &$\chi_r^{2,90}$ & Median $\chi_r^2$ & $E$(eV)\\
        \hline
        \HI & 17750 & 12.7 & 3.5 & 0.7 & 13.60\\
        \MgII & 5306 & 11.5 & 39.8 & 1.0 & 15.04\\
        \CII & 11062 & 12.8 & 15.8 & 1.3 & 24.38\\
        \SiIII & 14119 & 11.7 & 35.5 & 1.9 & 33.49\\
        \CIV & 17463 & 12.8 & 6.3 & 1.2 & 64.49\\
        \OVI & 17463 & 13.2 & 4.0 & 1.2 & 138.12\\
        \hline
    \end{tabular}
    \caption{Absorber sample properties for the RF models: the number of absorbers below the $\chi_r^2$ limit for each species; the column density completeness limit; the $\chi_r^2$ below which we recover 90\% of the total EW; the median $\chi_r^2$ of all absorbers; the excitation energy of the species.}
    \label{tab:absorber_sample_ml}
\end{table*}

In this work we use the sample of $z=0$ absorbers from our investigation into the physical conditions of absorbing halo gas in \citet{appleby_2023}. Here we summarise the procedure for generating the absorber sample. We select a sample of central galaxies within the fiducial \simba\ volume that evenly sample a range of global galaxy properties. \response{Central galaxies are defined as the one with the highest stellar mass among those in the halo; in practice, they are mostly within $\la 0.1r_{200}$ from the halo centre.}
The galaxies fall into three categories based on their star formation rates: star forming, green valley, and quenched. We define star forming galaxies as with ${\rm log (sSFR/Gyr}^{-1}) > -1.8 +0.3z$ for consistency with previous work with the \simba\ simulation \citep[e.g.][]{thomas_2019}, define green valley galaxies as within 1 dex below the star forming galaxy threshold, and define quenched galaxies as those having zero star formation. We further define six $\mstar$ bins of width 0.25 dex, with a minimum of $\mstar > 10^{10}\msolar$ to ensure well-resolved systems. In \cite{appleby_2023} we selected 12 galaxies from each of the 18 $\mstar-{\rm SFR}$ bin. Here we select a further 12 galaxies in each to double the underlying galaxy sample (except in the highest mass star forming and green valley bins, which have only 23 and 8 galaxies respectively) to increase the sample available for training a machine learning mapping.

For each central galaxy, we generate synthetic line of sight (LOS) absorption spectra through the simulation volume at a range of $r_{200}$-normalised impact parameters ($r_\perp$), probing both the inner and outer CGM ($r_\perp / r_{200}$ = 0.25, 0.5, 0.75, 1.0, 1.25). In addition, for each $r_\perp$, we select 8 equally-spaced LOS in a circle around the galaxy. Thus, for each galaxy in our sample we generate 40 LOS spectra for each of the following ions, selected to probe a range of excitation energies: \HI\ 1215\AA, \MgII\ 2796\AA, \CII\ 1334\AA, \SiIII\ 1206\AA, \CIV\ 1548\AA\ and \OVI\ 1031\AA.  This results in a total sample of \response{16600} lines of sight.

The spectra are generated along the $z$-axis of the simulation using the \pygad\ analysis package \citep{rottgers_2020}; the procedure is as follows. Gas elements whose smoothing lengths intersect with the LOS are identified and their ionisation fractions obtained, using look up tables that are generated with version 17.01 of the \cloudy\ cloud simulation code \citep{ferland_2017} using Cloudy Cooling Tools\footnote{\url{https://github.com/brittonsmith/cloudy\_cooling\_tools}}. We assume a spatially uniform \cite{faucher-giguere_2020} photoionising UV background spectrum, since it was shown in \citet{christiansen_2020} to provide the best match to low-redshift \lya\ absorption. Self-shielding for \HI\ is applied during the simulation run, but for generating the metal lines we employ the \cite{rahmati_2013a} prescription to attenuate the ionising background strength based on the local density.

Ion densities for each gas element are obtained by multiplying the gas densities by each species' ionisation fractions. The mass fractions of each element are individually tracked within \simba, based on yields from Type II and Ia supernovae and stellar evolution. Metals are carried out into the CGM primarily by stellar feedback processes, since winds are mass and metal-loaded \citep{appleby_2021}. The ion densities are smoothed along the LOS into pixels of width $2.5{\rm km\ s}^{-1}$, using the same spline kernel used in the \gizmo simulation code and the gas elements' individual smoothing lengths and metal masses (for metal lines). Optical depths are then computed from the column densities at a pixel scale, using the oscillator strength for each species. We exclude wind particles since those gas elements are hydrodynamically decoupled from the surrounding gas, which represent a very small fraction of the CGM mass~\citep{appleby_2021}. \pygad\ also computes column density-weighted physical density, temperature, metallicity, and peculiar velocity in the same manner within the LOS pixels. 

We identify regions of absorption within a $\pm 600 \kms$ window centered on the galaxy by computing the detection significance ratio of each pixel, defined as the Gaussian-smoothed flux equivalent width (EW) divided by the Gaussian-smoothed noise EW. Regions are identified as contiguous intervals where the flux drops below the level of the continuum with an overall significance ratio of $>4\sigma$, ensuring that the edges of the regions begin at the continuum and merging nearby regions within 2 pixels of one another.

We fit a superposition of Voigt profiles to each absorption region in order to extract the absorption line observables: the column density $N$, the Doppler $b$ parameter, the wavelength (or velocity) location along the LOS, and the EW. 
For the fitting, absorption lines are added to the model fit one at a time, with initial guesses for the line parameters that depend on whether or not the absorption is saturated. For non-saturated absorption, the line is placed at the position of lowest flux, and the initial $N$ and $b$ is based on the depth and velocity width of the local flux minimum. For saturated absorption, the line is placed in the middle of the saturated trough, and the initial $N$ and $b$ are chosen from a coarse grid in order to minimise the \response{reduced chi-square ($\chi_r^2$), computed assuming a signal-to-noise per pixel of 30}.  This procedure broadly follows that in AutoVP~\citep{dave_1997}.
The best-fit Voigt parameters that minimise $\chi_r^2$ are then found using the {\tt scipy.optimize} subpackage\footnote{\url{https://docs.scipy.org/doc/scipy/reference/optimize.html}}. Loose prior bounds on $N$ and $b$ are set based on typical \HI and metal line column densities and thermal line widths from $10^4-10^7$K.

If the fit has $\chi_r^2 < 2.5$ then the model is accepted; otherwise we identify the next strongest area of absorption by subtracting the model from the data and place a line at the residual minimum. We repeat the process until an acceptable model is found, up to a maximum of 10 absorption lines per region. Each line must improve the $\chi_r^2$ of the model by at least 5\%; the process is halted if 2 consecutive additional lines do not improve the $\chi_r^2$ by at least this margin. If after 10 lines an acceptable model is not found then we adopt the model with the number of lines that performed best. We again check that each line improves the $\chi_r^2$ of the model by iteratively recomputing the $\chi_r^2$ with each line removed; if the $\chi_r^2$ acceptance threshold is reached, or the $\chi_r^2$ increases by less than 5\% then the line is removed from the solution.  In this way we attempt to obtain a satisfactory fit with the fewest number of absorption lines.

\response{\pygad not only produces optical depths at each pixel, but also outputs the optical depth-weighted density, temperature, and metallicity (in the relevant element, or the total metallicity in the case of H), as described in \citet{appleby_2023}.  This enables us to assign physical properties to absorption features.  In our case, we assign to each absorber the physical properties associated with the pixel closest to its Voigt profile fitted line centre.  We prefer this to a weighting or interpolation scheme, because if there are significant variations in the physical properties between pixels then interpolation can yield values that are less physically meaningful~\citep{ford_2014}.  Nonetheless, this is an inherently approximate procedure, which is only exact in the unphysical situation that each  Voigt profile represents a distinct uniform cloud of absorbing gas.  We further note that sub-resolution phenomena such as small-scale turbulence is not accounted for, which could additionally blur the relationship between the absorber temperature and the Voigt profile line width.}

The galaxy selection, spectrum generation and LOS fitting pipeline results in our sample of absorbers. We find that adopting the same strict $\chi_r^2$ limit for all ions results in an incomplete sample. As such we compute the EW directly for each LOS and adopt an upper $\chi_r^2$ threshold for each ion such that we recover 90\% of the total EW across all LOS for each species. The sample size and $\chi_r^2$ upper limits ($\chi_{r,90}^2$) for each ion are given in Table \ref{tab:absorber_sample_ml}. In practice the typical $\chi_r^2$ for a given region is much lower than these upper limits; the median $\chi_r^2$ of absorption lines in our sample is also shown in Table \ref{tab:absorber_sample_ml}. We also adopt the column density completeness limits from \cite{appleby_2023}, which are computed by fitting the power law portion of the column density distribution function (CDDF) for each ion and identifying where the CDDF falls below 50\% of the expectation at low column densities. The completeness limits are given in Table \ref{tab:absorber_sample_ml}.  We note that routines to do the spectrum generation, absorption region identification, and Voigt profile fitting are all contained with the publicly-available \pygad\ package \citep{rottgers_2020}.

\section{Random Forest Methods}\label{sec:random_forest}

\subsection{Random Forest Regression }

Random Forest (RF) regression \citep{breiman_2001} is a supervised, decision tree-based, ensemble method of machine learning. The term `ensemble method' refers to the process of combining predictions from several machine learning runs (in this case, individual decision trees) in order to more accurately predict the output. Decision trees work in a top-down manner, in which the best split for the data is found by minimising a cost function. They have the advantage of being easy to interpret and have low bias in their predictions for the training data. However, individual decision trees are prone to over-fitting to the training data, hence their predictions for new data have high variance.

The RF algorithm counteracts this effect by constructing many decision trees, each trained on a subset of the data. Random forests may be used for both classification and regression problems; in this work we use RF in its regression mode to deal with our continuous target predictors. The training data subsets are randomly chosen with replacement, and their outputs averaged for an overall prediction in a process known as bootstrap aggregation \citep[`bagging', see][]{breiman_1996}. In this way, RF models retain the low bias of a decision tree, while also minimising the variance on predictions for new data. Training a single decision tree is considerably faster, however such models are less reliable, particularly when trained on non-linear data (such as the absorber data used here). In this work, we use the {\tt Scikit-Learn} \citep{pedregosa_2011} module's RF implementation, {\tt RandomForestRegressor}.

RF models are widely used in a range of astronomical applications, and have been remarkably successful given the relative simplicity of the approach. The advantage of RF models over other methods (for example Neural Network based algorithms) is in the interpretability of the output models, as they indicate the relative importance of the input variables in reaching a prediction.
In galaxy formation, RFs (and related tree-based methods) have been widely used for regression problems using both simulation and observational data, for example in predicting the properties of large scale structure \citep{lucie-smith_2018, lovell_2022, li_2022} and the properties of galaxies and haloes \citep{ucci_2017, nadler_2018, rafieferantsoa_2018, cohn_2020, moews_2021, mucesh_2021, delgado_2022, mcgibbon_2022}.

\subsection{Input features and target predictors}

For each of the ions in our selection, we train a RF model on the dataset of simulated CGM absorbers to predict their underlying physical gas conditions. We do this separately for each of the 6 ions we consider, such that the usefulness of this pipeline is not contingent on having line information simultaneously for all 6 ions. We exclude absorbers where the quality of the Voigt profile fit is low (i.e. the fit has a $\chi^2_r$ above the acceptable threshold for that ion) and the column density is below the completeness limit. 

For each ion, we use the same set of input features and target predictors. The input features are chosen from among the properties of the CGM absorbers and their central galaxies. Included features which describe the absorbers themselves are: the column density ($N$), the equivalent width (EW), the linewidth ($b$), the velocity separation from the host galaxy (${\rm d}v$), and the impact parameter, expressed as a fraction of halo virial radius ($f_{r200} = r_\perp / r_{200}$). \response{In principle, the EW information is fully contained within $N$ and $b$, but we provide it separately since sometimes combinations of parameters can be easier for the ML to utilise, and also so we can compare between $N$ and $EW$ to see which measure of overall absorption is more important.} 

Properties of the central galaxy that are included as input features are the stellar mass ($\mstar$), the specific star formation rate (sSFR), and the fraction of kinetic energy contained in rotation \citep[$\kappa_{\rm rot}$,][]{sales_2012}, which \citet{kraljic_2020b} found is a reasonable proxy for visual morphology.

From these 8 input features we predict 3 target gas predictors: the  overdensity ($\delta = \rho / \bar{\rho}_m$), temperature ($T$) and metallicity ($Z$). Each of these is a column density-weighted average at the nearest LOS pixel to the absorber, computed at the time of spectral generation and binned along the LOS (see \S\ref{sec:absorbers}). 

\subsection{Training}

Each of the features is transformed into log space; \cite{jo_kim_2019} showed that transforming quantities into log space improves the accuracy of machine learning predictions for astronomy problems, owing to the wide range of physical scales present in astronomical data. Exceptions to this are ${\rm d}v$, $f_{r200}$ and $\kappa_{\rm rot}$; ${\rm d}v$ and $\kappa_{\rm rot}$ have nearly uniform intrinsic distributions, while $f_{r200}$ consists of 5 specific values due to our choices of LOS. In addition, we standardize the input and output data by subtracting the mean of the distribution and scaling the variance to unity in each case. 
\response{We deal with zeros in our dataset by setting them to a small non-zero value; in practice this can only be the case for sSFR, whereby we assign sSFR = $10^{-14}\msolar {\rm yr}^{-1}$.}
For each ion, we divide the absorber dataset into 80\% training data used to build the RF model, and 20\% test data used to evaluate the performance of the model. Where multiple absorbers arise from the same LOS these can be separated into the training and test datasets; this mitigates over-fitting in the model due to galaxy or LOS properties.

We train the RF model separately for each target feature, as we find that this improves the accuracy of the prediction. We separately tune the hyperparameters of each RF model to optimize the model accuracy, using {\tt Scikit-Learn}'s {\tt GridSearchCV} method to perform an exhaustive grid search over hyperparameter space. The hyperparameters are the number of trees, the minimum number of data points required in order to split the data, and the minimum number of data points in each resulting split. For each set of hyperparameters, a $k$-fold cross validation is performed with $k = 5$, in which the training data is split into $k$ `folds', and $k$ RF models are iteratively  constructed using $k-1$ folds of the data; the overall score for each set of hyperparameters is the average of each of the $k$ RF models. The coefficient of determination $R^2$ is used internally \response{at this hyperparameter tuning stage} to evaluate the performance of each model, given $n$ data points and true and predicted quantities $X_{\rm true}$ and $X_{\rm predicted}$:
\begin{equation}
    R^2 = 1 - \frac{\rm RSS}{\rm TSS} 
\end{equation}
where RSS is the residual sum of squares:
\begin{equation}
    {\rm RSS} = \sum\limits_{i=1}^n (X^i_{\rm true} - X^i_{\rm predicted})^2
\end{equation}
and TSS is the total sum of squares:
\begin{equation}
    {\rm TSS} = \sum\limits_{i=1}^n (X^i_{\rm true} - \langle X_{\rm true}\rangle)^2
\end{equation}
The mean squared error, MSE $= {\rm RSS} / n$, is the cost function used to determine the best decision tree splits. An MSE of zero represents a perfectly accurate prediction.

\section{Predictive accuracy}\label{sec:ml_accuracy}

\begin{figure*}
	\includegraphics[width=\textwidth]{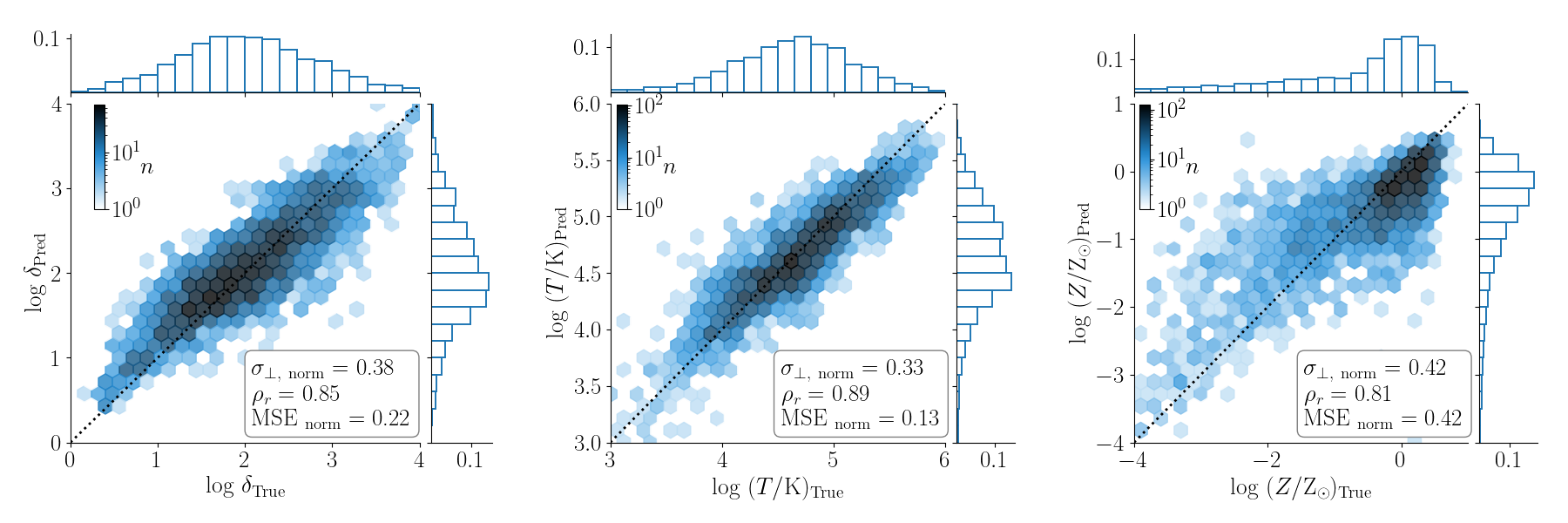}
	\vskip-0.1in
    \caption{Hexagonal joint histogram of the predicted \HI\ physical conditions from the RF mapping and the true \HI\ physical conditions, including only data in the test set. The number of data points in each bin is shown using colorbars. From left to right, the panels show overdensity, temperature and metallicity. The diagonal line represents the case where the RF model makes a perfect prediction. The accuracy of the predictions in each panel is summarised by the inset displaying the \response{normalised} transverse scatter $\sigma_{\perp,\ {\rm norm}}$, the correlation coefficient $\rho_r$ and the \response{normalised} mean square error, MSE$_{\rm\ norm}$. The 1D histograms of the true and predicted values are shown on the top and side of each panel, respectively. } 
    \label{fig:hi_joint}
\end{figure*}

Here we assess the performance of each of the RF models.
Figure \ref{fig:hi_joint} shows the test data RF predictions for the \HI\ absorber physical conditions ($\delta$, $T$ and $Z$) against the true values. The color scale of the hexagonal bins indicates the number of data points in each bin. The black diagonal dashed line represents the 1:1 case of a perfectly predicting model. In each panel the 1D histograms for the true and predicted values lie along the top and right, respectively. The accuracy of the model is summarised in each panel with three quantities: 1) \response{$\sigma_{\perp, {\rm norm}}$, the scatter perpendicular to the perfect 1:1 relation, normalised to the scatter in the true values; }2) the Pearson correlation coefficient $\rho_r$, given by:
\begin{equation}
    \rho_r = \frac{{\rm cov} (X_{\rm true}, X_{\rm predicted})}{\sigma_{X{\rm true}} \sigma_{X{\rm predicted}}},
\end{equation}
where $\sigma_{X i}$ is the standard deviation of $X_i$; and 3) \response{MSE$_{\rm norm}$, the mean squared error normalised to the scatter in the true values}. High correlation is preferred, but does not necessarily indicate an accurate prediction as the outputs could have a systematic offset. \response{We normalise $\sigma_\perp$ and MSE in order to compare results between \HI and the metals lines, since \HI lines trace a wider range of physical conditions.}

Beginning with the predictions for \HI\ absorbers, density and temperature are well-predicted by the ML model. True values are highly correlated with the predictions, and the model predictions have low scatter and error. Density and temperature are physically correlated with one another and have Gaussian distributions. Of the two, temperature (\response{$\sigma_{\perp,\ {\rm norm}} = 0.33$ dex, MSE$_{\rm\ norm}$ = 0.13)} is predicted more accurately than overdensity \response{($\sigma_{\perp,\ {\rm norm}}= 0.38$ dex, MSE$_{\rm\ norm}$ = 0.22). }
The RF models for HI density and temperature perform particularly well considering the models' relative simplicity (compared with e.g. a NN-based model). Aside from transforming the features into log space and using the $k$-fold hyperparameter cross validation, the model has not been extensively tuned by hand. As such, these results represent a basic model which demonstrate the capability of RF models to predict gas conditions, which could be improved upon with further tuning. We have also explored alternative ML approaches such as NNs and CNNs, and found that such models do not offer a substantial improvement in terms of predictive accuracy and take considerably longer to run. This has lead us to favour the RF model for its simplicity, speed, and the degree of interpretability in the form of feature `importances' (see \S\ref{sec:features}).

The predictions for HI metallicities are less accurate \response{($\sigma_{\perp,\ {\rm norm}} = 0.42$ dex, MSE$_{\rm\ norm}$ = 0.42)}. In general, points with ${\rm log} Z/Z_{\odot} <-1$ are overpredicted, while points with ${\rm log} Z/Z_{\odot} >-1$ are underpredicted. This points to the general tendency of our ML models to output a narrower predicted distribution than in the input dataset (this behaviour is also seen to a lesser extent in the density and temperature predictions). This means that the tails of the original distributions are not well captured in the ML model, perhaps as a result of sparse training data at the extremes. Perhaps the poor prediction is unsurprising since HI absorption is not $Z$-dependent, unlike metal lines which by necessity arise from metal-enriched gas.  Therefore it was not obvious that any relationship between \HI\ absorption and metallicity could have been learned from the data. The learned mapping in the metallicity RF model likely arises from the provided galaxy properties and \HI absorption strength; we will explore the input feature importance later (\S\ref{sec:features}).

\begin{figure*}
	\includegraphics[width=\textwidth]{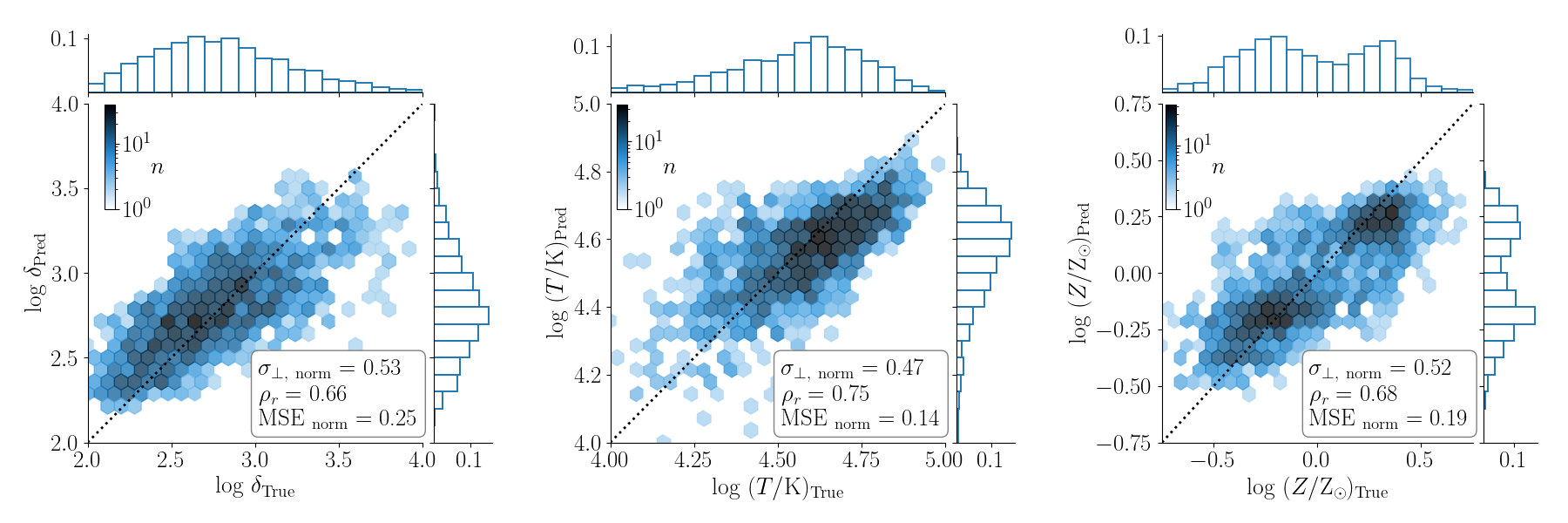}
	\vskip-0.1in
    \caption{As in Figure \ref{fig:hi_joint}, showing the predictions and true values for \CII\ absorbers.} 
    \label{fig:cii_joint}
\end{figure*}

\begin{figure*}
	\includegraphics[width=\textwidth]{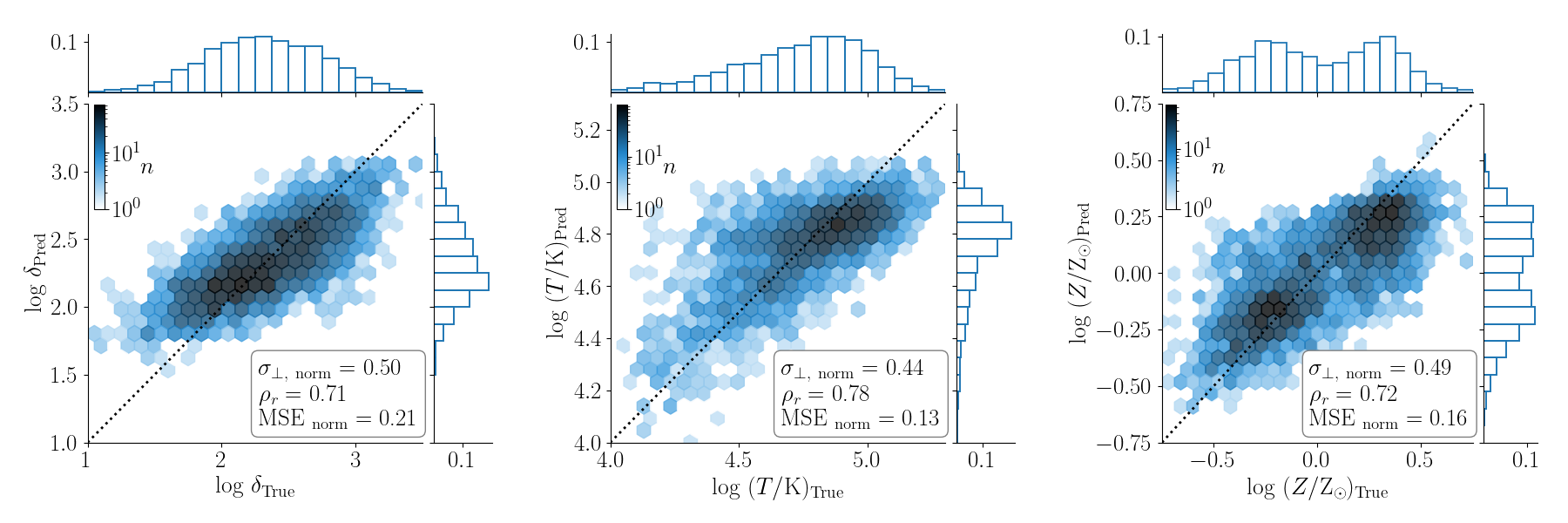}
	\vskip-0.1in
    \caption{As in Figure \ref{fig:hi_joint}, showing the predictions and true values for \CIV\ absorbers.} 
    \label{fig:civ_joint}
\end{figure*}

The metal line absorber physical conditions are also reasonably well predicted. Figures \ref{fig:cii_joint} and \ref{fig:civ_joint} show the performance of the RF models for predicting \CII and \CIV absorber conditions, using the same plot structure as above. The performance for \MgII, \SiIII and \OVI absorbers are shown in Appendix \ref{sec:additional_metals}. The RF models perform similarly well among all the metal lines, with the same tendency to predict a more concentrated distribution of values than in the original data. In general, for each metal line the predictions are less well correlated with the truth values than for \HI; metal line absorber $\delta, T$ and $Z$ have median correlation coefficients of \response{$\rho_r = 0.68, 0.71, 0.68$}, respectively, compared with $\rho_r = 0.85, 0.88, 0.81$ for \HI. 
\response{In addition, the scatter is higher in general for the metal line RF models, with median $\sigma_{\perp,\ {\rm norm}} = 0.52, 0.44, 0.52$ for $\delta, T$ and $Z$, compared with $\sigma_{\perp,\ {\rm norm}} = 0.38, 0.33, 0.42$ for \HI.
However, the errors in the metal line predictions are comparable with those for \HI: median MSE$_{\rm\ norm} = 0.24, 0.13, 0.18$ for metal lines, compared with MSE$_{\rm\ norm} = 0.22, 0.13, 0.42$ for \HI.
}
Overall the RF models give reasonable predictions for the physical conditions, and again were not extensively tuned to achieve this. 

An interesting feature of the original absorber dataset is bimodal metallicity distributions at ${\rm log}Z/{\rm Z}_\odot \sim -0.25\ {\rm and}\ 0.25$, which have not been reported in earlier \simba\ CGM work. The bimodality is apparent in every metal line apart from \MgII, and is broadly reproduced by the RF models.
Populations of absorbers in the cool CGM of low redshift Lyman Limit Systems (LLSs) have also been observed to have bimodal metallicity distributions, with both metal-poor and metal-rich absorbers \citep[albeit shifted to lower metallicities,][]{lehner_2013, lehner_2018, lehner_2019, wotta_2016, wotta_2019, berg_2023}, suggesting multiple origins for the cool CGM gas, although the metallicities of the observed metal-poor absorbers are much lower than that seen in \simba. In future work we will investigate the origin of the bimodal absorber metallicity distribution in \simba.

\section{Feature importance}\label{sec:features}

In this section, we seek insights into the physical origin of the ML-probed correlations by assessing which input features are most useful in predicting the physical conditions.

\subsection{RF model importance}\label{sec:importance}

An advantage of the RF method over other ML algorithms (such as neural networks) is it allows some degree of interpretability in the form of the `importance' of each feature, which arise `for free' from the structure of the RF model. For an individual decision tree, a feature's importance is computed from the number of times it is used to split the data and how close to the top of the tree the splits are. For an RF model, the importances are the normalised average over all decision trees. However, importance metrics are biased if the input features are highly correlated with one another \citep[][]{strobl_2007, strobl_2008} and so they should be treated with caution.  Thus we prefer not to use the importance directly reported by RF, but instead compute it more empirically.

To do so, we determine each input feature's importance by iteratively building the RF model and removing each of the features in turn, using the same optimized hyperparameters as in the full feature model. We then retrieve the importance of the remaining features for each model. 
This process determines whether a feature is genuinely important, or merely defined as such through a fluke of feature combinations. When the most important features are removed, identifying which features take its place as the most important gives an indication of what the RF model is learning. 

\begin{figure*}
	\includegraphics[width=\textwidth]{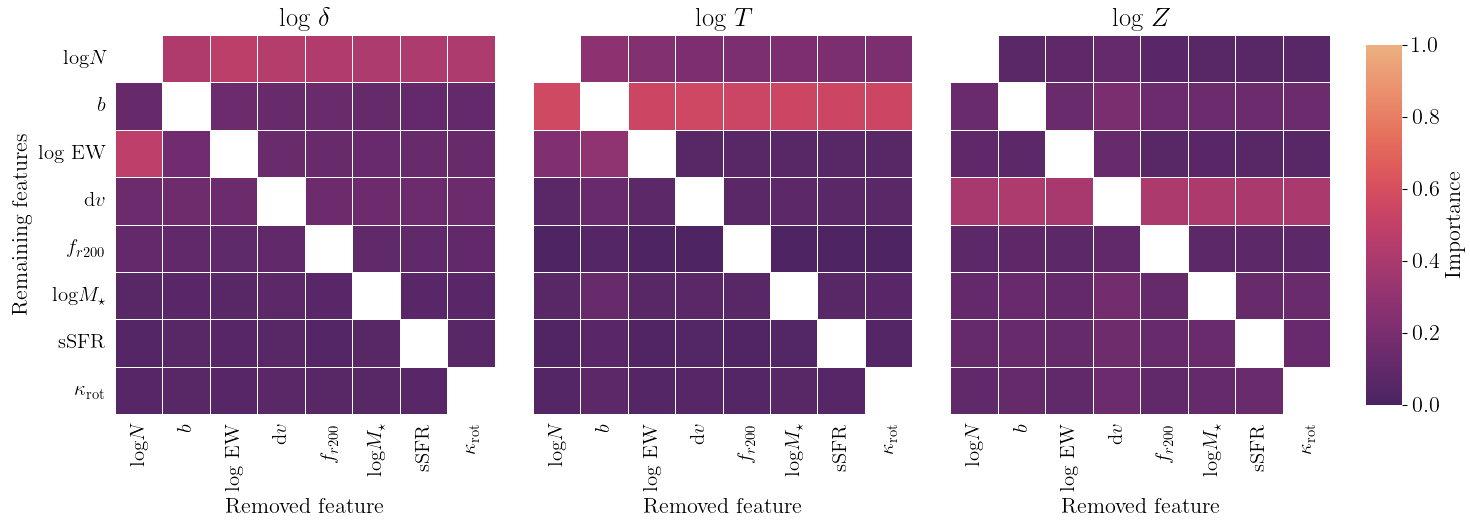}
	\vskip-0.1in
    \caption{Importance values in predicting \HI\ physical conditions for each remaining input feature, against the input feature removed from the training data. From left to right, the target predictors are $\delta, T$ and $Z$.} 
    \label{fig:hi_importance}
\end{figure*}

Figure \ref{fig:hi_importance} shows the feature importance values for predicting \HI\ absorber conditions, against the feature removed from the training data. 
For predicting overdensity, $N$ is most important feature. When column density is removed, the most important feature is EW; $N$ and EW are correlated with one another and both are correlated with physical density. When predicting temperature, $b$ is the most important feature since the linewidths of individual absorbers in the original spectrum are set in partly by thermal Doppler broadening (with the additional effect of bulk gas motions). 
When predicting metallicity, the velocity separation is the most important feature. It is not intuitively obvious why this is the case; perhaps due to a dependence on halo velocity dispersion, which is correlated with $\mstar$ and thus the metallicity of the host galaxy that is predominantly responsible for enriching its CGM.

\begin{figure*}
	\includegraphics[width=\textwidth]{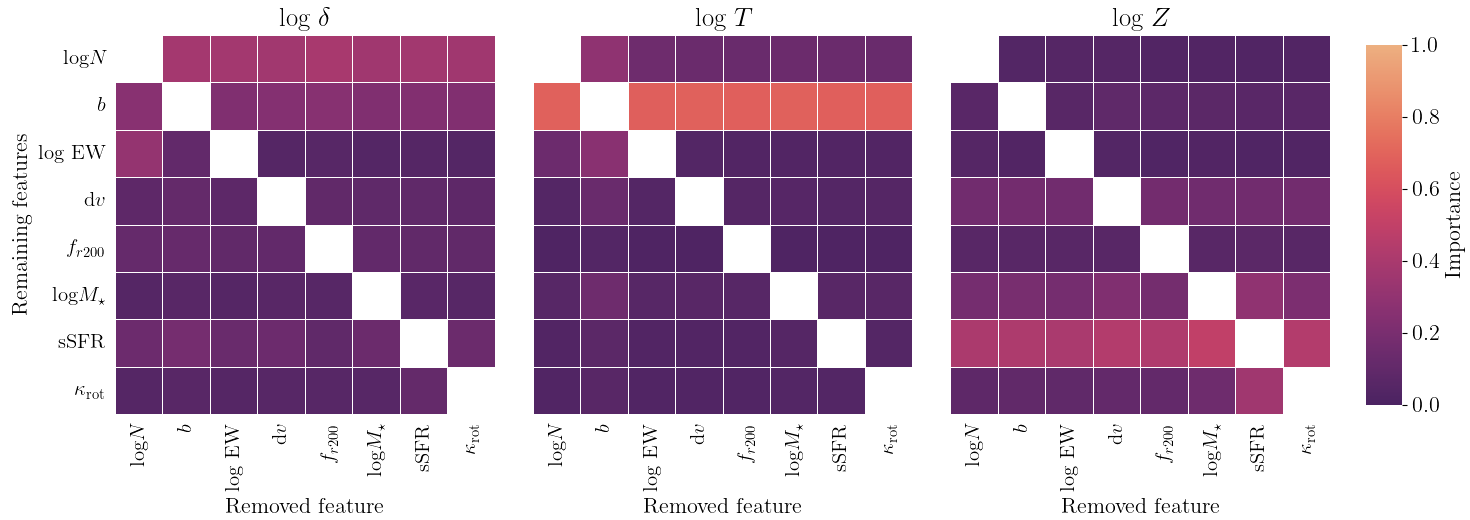}
	\vskip-0.1in
    \caption{Feature importance values as in Figure \ref{fig:hi_importance}, for CII absorbers.}
    \label{fig:cii_importance}
\end{figure*}

\begin{figure*}
	\includegraphics[width=\textwidth]{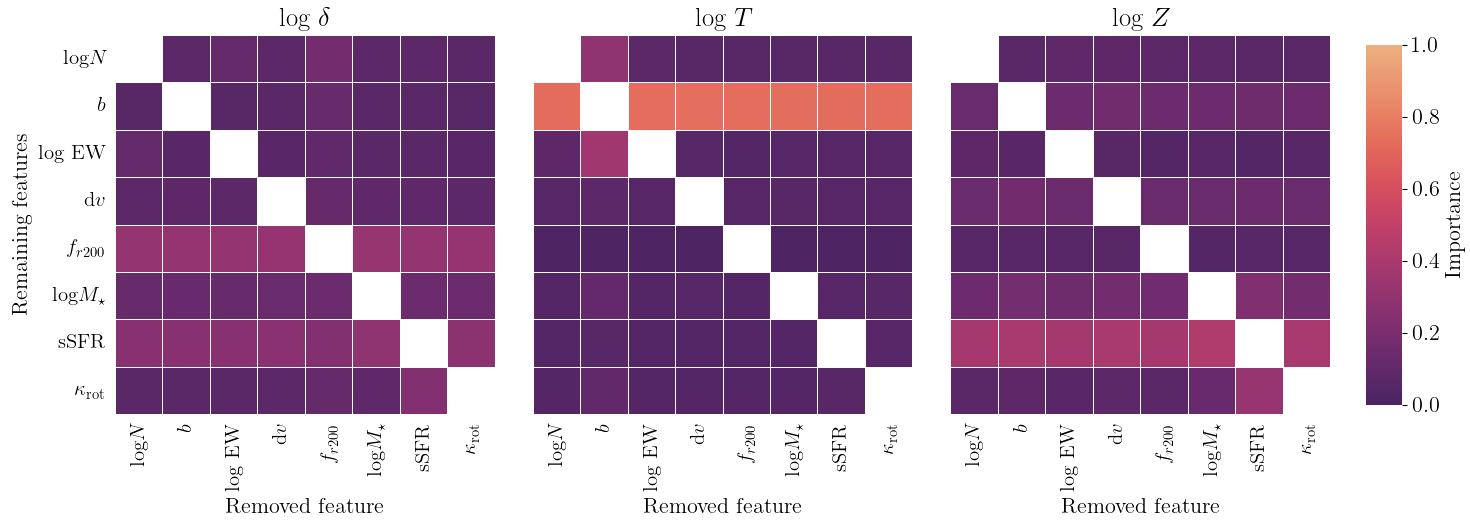}
	\vskip-0.1in
    \caption{Feature importance values as in Figure \ref{fig:hi_importance}, for CIV absorbers.} 
    \label{fig:civ_importance}
\end{figure*}

Figures \ref{fig:cii_importance} and \ref{fig:civ_importance} likewise show the feature importance values for predicting \CII\ and \CIV\ absorber properties. We have examined feature importance for all metals and found that these are representative cases. For the low ion \CII, the feature importance rankings for $\delta$ and $T$ are similar to that of \HI. There is a slightly reduced relative importance of $N$ in predicting $\delta$ in favour of $b$ ($N$ and $b$ are correlated features due to their underlying dependence on $\delta$ and $T$). For both ions, the importance of $b$ in predicting $T$ is enhanced compared with \HI. In contrast to \HI, the most important feature for predicting $Z$ for metal lines is sSFR; when sSFR is removed, the RF model learns from $\mstar$ and $\kappa_{\rm rot}$ instead, indicating that the RF model predicts $Z$ from the galaxy properties. The picture is similar for the high ion \CIV, except that in predicting $\delta$ the most important features are instead $f_{r200}$ and sSFR. $f_{r200}$ is perhaps less useful for \CII\ since most of the low ion absorption arises from the inner CGM \citep[][]{appleby_2023}. Galaxies with high star formation have denser gas in their CGM, although this is also the case for \HI\ and \CII, so it is unclear why sSFR specifically is an important feature for \CIV\ absorbers.



\subsection{Change in predictive accuracy}

Arguably the most meaningful measure of `importance' to the model is in which features add the most useful information in terms of predictive accuracy. 
We assess this by iteratively removing each of the features in turn from the training data, and running the RF model as before (with the hyperparameters optimized for the full feature case). In contrast with \S\ref{sec:importance}, we now compute the scatter $\sigma_{\perp,\ {\rm norm}}$ for each new model; if the quality of the predictions are significantly degraded in the absence of a particular feature, this necessarily indicates that this feature encodes crucial information about the physical conditions.

\begin{figure*}
	\includegraphics[width=\textwidth]{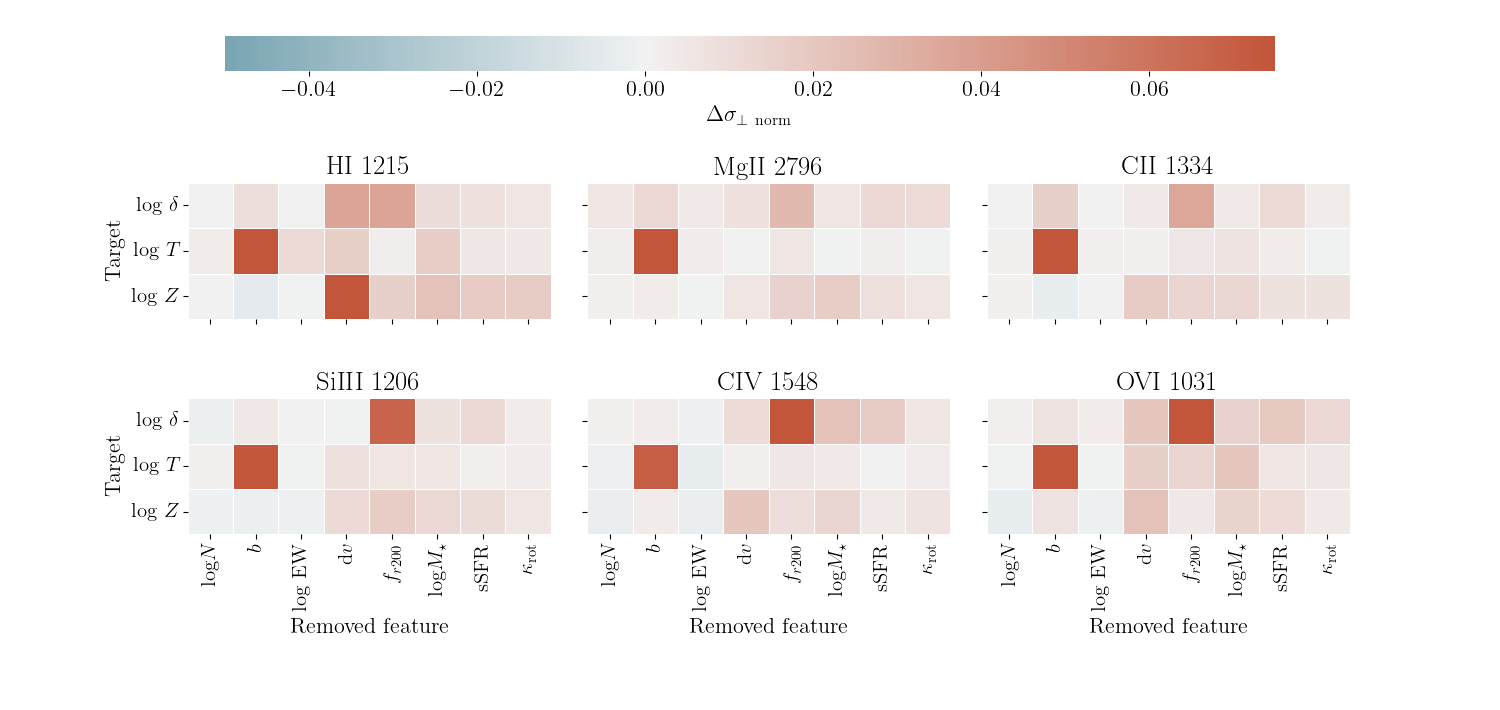}
	\vskip-0.1in
    \caption{The change in $\sigma_{\perp,\ {\rm norm}}$ of the RF models when removing each feature iteratively. Each of the panels shows results for a different ion; the three rows of each panel represent results for each of $\delta, T$ and Z.} 
    \label{fig:delta_scatter}
\end{figure*}

Figure \ref{fig:delta_scatter} shows the change in $\sigma_{\perp,\ {\rm norm}}$ resulting from the removal of each input feature in predicting the physical conditions, where a positive change indicates an increase in scatter. Each of the six panels shows the models for a different ion; the rows within each panel show the models for each of $\delta, T$ and $Z$. The upper plots show \HI\ (left), \MgII\ (middle) and \CII\ (right) absorber models, while the lower plots show \SiIII\ (left), \CIV\ (middle) and \OVI\ (right).

In contrast to the feature importance values, the largest increase in scatter when predicting \HI\ absorber overdensity comes from removing ${\rm d}v$ and $f_{r200}$. The loss of accuracy from removing $f_{r200}$ suggests that the RF model is learning the radial density profile; this is also the case for the metal line $\delta$ predictions. It is less clear why ${\rm d}v$ is necessary for an accurate prediction, since halo absorbers can appear at any velocity separation depending on their kinematics. For all ions, removing $b$ causes an increase in scatter in $T$ predictions, confirming that the high feature importance of $b$ reflects the genuine physical relationship with temperature.

For \HI, the metallicity predictions are degraded by removing ${\rm d}v,\ f_{r200}$ or any galaxy property; interestingly the model accuracy improves when absorption-related features are removed. In other words, gas metallicity in the CGM of a given galaxy can be predicted with reasonable accuracy from only LOS and radial absorber position. The metal lines broadly show the same changes in scatter with removed features, although the changes to the model accuracy are more marginal.

\response{In some cases, $\sigma_{\perp,\ {\rm norm}}$ actually reduces with a given input feature removed.  Ideally, this should never happen, as more information should always result in a better fit.  Thus this indicates that perhaps there is some slight overfitting by the ML algorithm, or else there is some stochasticity in fitting process.  The fact that these reductions are generally quite small even when present, much smaller than the typical increases, suggests that this is not a significant issue in the pipeline.}

\section{Phase space}\label{sec:phase_space}

Having examined the predictive accuracy and inner workings of the RF models on individual properties, we now ask whether the RF models can reproduce the two-dimensional $(\delta,T)$ phase space structure of the absorbers. Although the RF models can separately predict $\delta$ and $T$, this does not guarantee that they reproduce the relationship between these quantities - particularly since the models are trained separately for $\delta$ and $T$, so each RF model has no knowledge of the other target quantities.

\begin{figure*}
	\includegraphics[width=\textwidth]{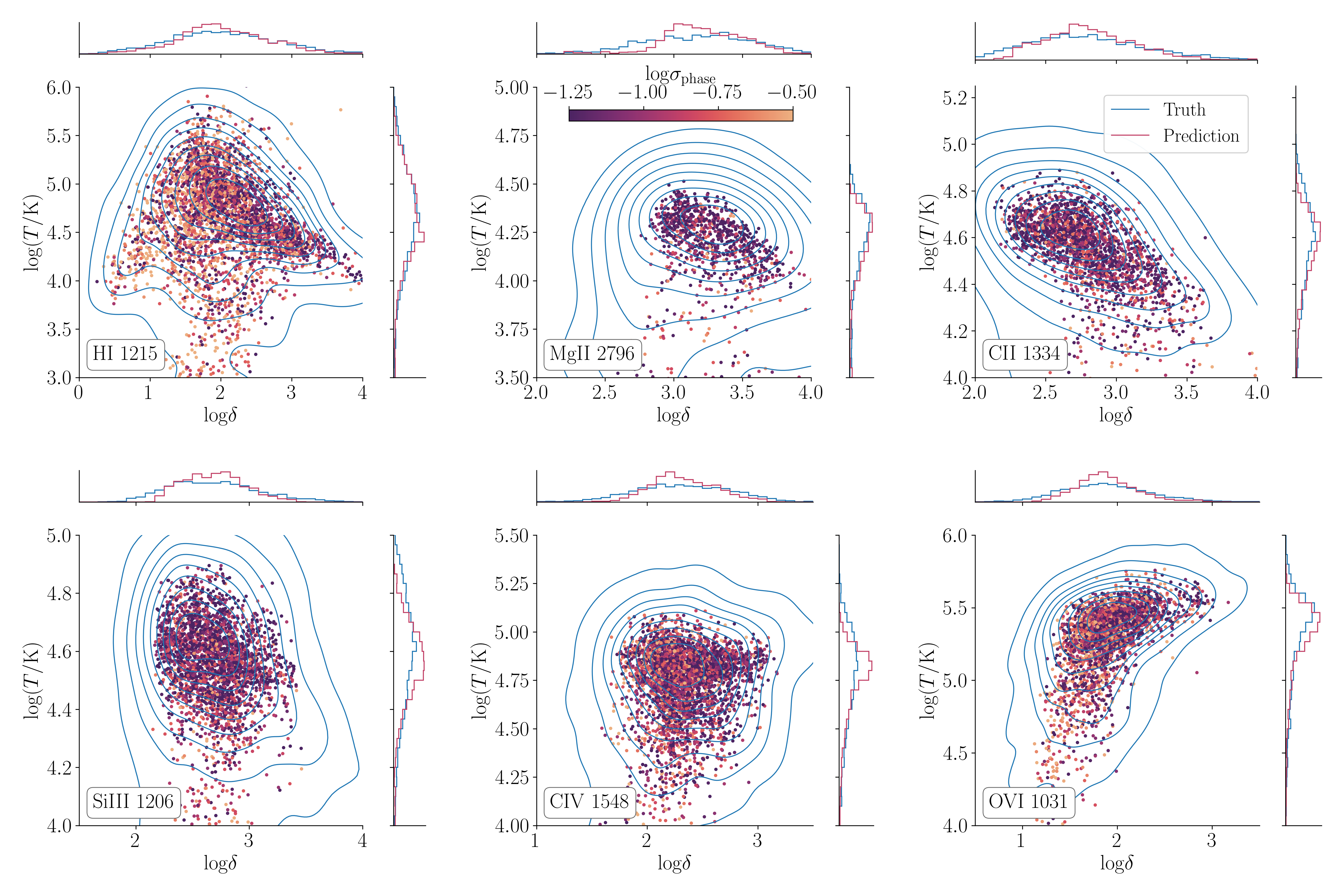}
	\vskip-0.1in
    \caption{Predicted temperature against predicted overdensity for each of the 6 ions we consider, coloured by overall phase space fractional error. The 1D truth (blue curve) and predicted (pink curve) distributions are shown along the top and right of each panel. The contours show the true distribution in phase space for the test dataset. The limits of the plots differ for each ion.} 
    \label{fig:phase_space}
\end{figure*}

Figure \ref{fig:phase_space} shows the predicted temperature against predicted overdensity for the 6 species we consider. The distributions for the truth and predicted data are shown along the top and right hand side of each panel. Note that the plot limits are different for each ion. The points are colour coded by the fractional distance from the true value in $\delta-T$ phase space:
\begin{equation}
    \sigma_{\rm phase} = \sqrt{\Big(\frac{\delta_{true} - \delta_{\rm predict}}{\delta_{\rm true}}\Big)^2 + \Big(\frac{T_{true} - T_{\rm predict}}{T_{\rm true}}\Big)^2}
\end{equation}
\response{By this metric, a higher $\sigma_{\rm phase}$ (orange in the colour map) represents a larger distance from te truth value.} The colour scale indicates that in general the predicted points lie near their truth values in phase space; the metal line absorbers in particular have a low displacement.
The contours show the true distribution in phase space for the test dataset. The upper plots show \HI\ (left), \MgII\ (middle) and \CII\ (right) absorbers, while the lower plots show \SiIII\ (left), \CIV\ (middle) and \OVI\ (right). 
The original structure in phase space between overdensity and temperature is reproduced well by the RF models for each of the ions we consider, a success of the ML approach which (since there was no specific tuning of the model to achieve this) arises because temperature information is encoded in the overdensity data and vice versa. This is an important test of the RF models which verifies that accurate predictions can be made for multiple physical properties per observation.

That said, although the RF models succeed in predicting the phase space structure, the predictions are in general too concentrated near the mean of the data. By comparing the predicted distribution with the contours from the original data, it is clear that the predictions ought to be more spread in phase space. This appears to be a generic feature of the RF models, which is also apparent in the 1D distributions of each feature - in general, the RF models produce predicted distributions that are too concentrated towards the mean. As mentioned in \S\ref{sec:ml_accuracy}, this likely arises from sparse training data at the extremes.
Thus, the predicted distributions do not capture the important information described by the intrinsic scatter in the original data, which biases the usefulness of these models for observational analysis.
Therefore some additional step beyond the basic ML model is required in order to capture the full structure in phase space. Employing an oversampling technique (such as Synthetic Minority Over-Sampling Technique for Regression with Gaussian Noise, SMOGN) can boost under-sampled regions of phase space and as such mitigate issues that arise as a result of imbalanced training datasets \citep{de_santi_2022}.

\begin{figure*}
	\includegraphics[width=\textwidth]{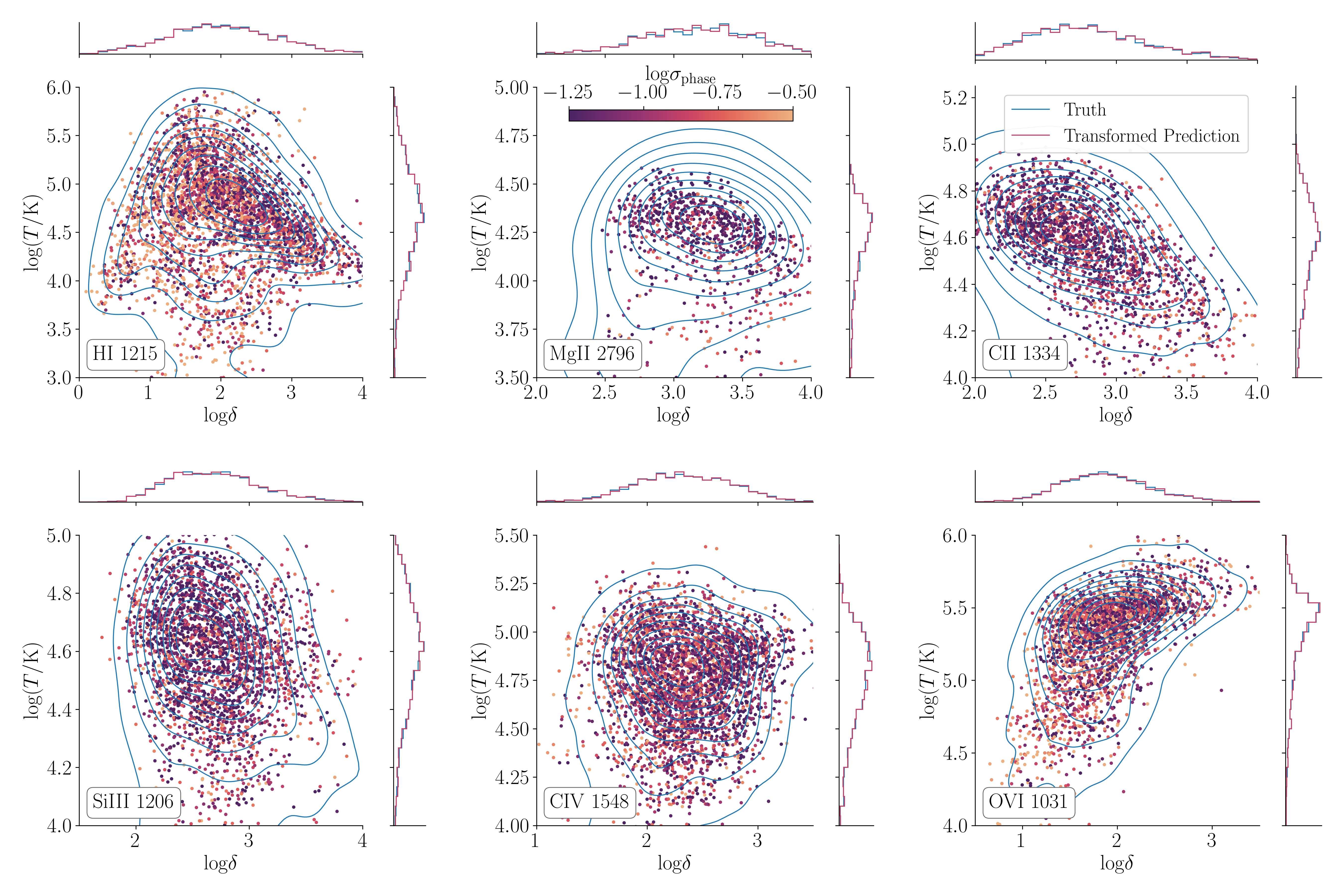}
	\vskip-0.1in
    \caption{Predicted temperature against predicted overdensity for each of the 6 ions we consider, mapped to the shape of the truth 1D distributions using a quantile transformer. Points are coloured by overall phase space fractional error. The 1D distributions for the truth data (blue curve) and transformed predictions (pink curve) are shown along the top and right of each panel. The contours show the true distribution in phase space for the test dataset.} 
    \label{fig:phase_space_trans}
\end{figure*}

In order to extend the RF network to also properly capture the full 2-D phase space distribution, we develop a new approach based on a normalising transform.  
By this, we mean that the predicted and truth data for each feature are mapped onto standard normal distributions, and then the predicted distribution is transformed back onto the shape of the truth data distribution.

To accomplish this we use the quantile transform non-parametric method implemented in {\tt Scikit-Learn} \citep{pedregosa_2011}, {\tt QuantileTransformer}. The method first maps the cumulative distribution of the data onto a standard Gaussian, and then computes the transformed values using a quantile function.
The function also provides the inverse mapping that transforms a distribution back into the original coordinates.
The inverse mapping for the truth data distribution (computed from the training dataset) is used to reconstruct the predicted distribution.
In this way, we can reproduce the larger variance in the truth data without assuming a shape for the predicted data.

Figure \ref{fig:phase_space_trans} shows the the predicted temperature against predicted overdensity, using the above normalising transform approach to map the shape of the truth data. This can be seen in the 1D distributions along the top and right hand side, which in most cases closely follow the truth data distributions. Crucially, the transformed data also retains the phase space structure of the original predictions. In addition, transforming the predictions onto the shape of the truth data results in no loss of accuracy for the predictions; the MSE$_{\rm\ norm}$, $\rho$ and $\sigma_{\perp,\ {\rm norm}}$ for the transformed test datasets are very similar to those for the original test dataset predictions. As such, this is our preferred method for reproducing the scatter in the truth CGM conditions data. 

\begin{figure*}
	\includegraphics[width=\textwidth]{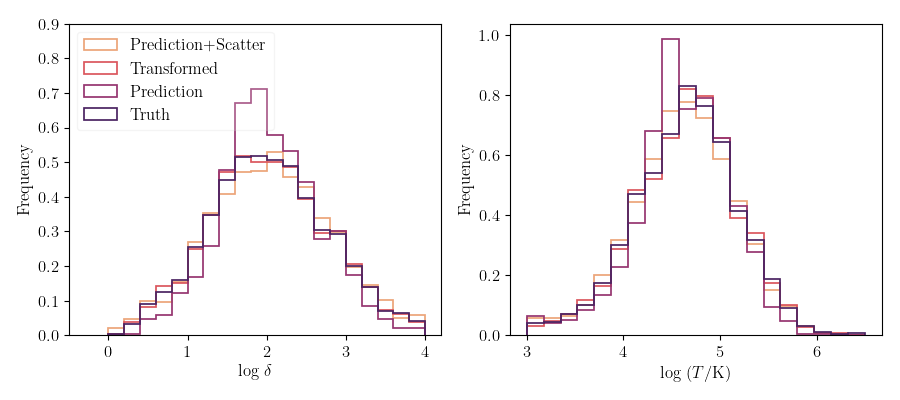}
	\vskip-0.1in
    \caption{Histograms showing the direct comparison between truth data (dark purple), the predictions (light purple), the transformed predictions (orange) and the predictions with added scatter (yellow) for \HI overdensities and temperatures. } 
    \label{fig:HI_scatter_hists}
\end{figure*}

We initially explored a simpler approach where we added scatter directly to the predicted data.
The results in phase space for the additional scatter approach is shown in Appendix B (Figure \ref{fig:phase_space_scatter}).  We found that this approach substantially washed out non-Gaussian structure in the predicted distribution, such as the anti-correlation between $\delta$ and $T$ at $\delta>10^3$. 
By instead using the normalising transform method, we are preserving these structures in phase space as much as possible.

Figure \ref{fig:HI_scatter_hists} directly compares the distributions for the truth data (dark purple), the predictions from the RF models (light purple), the transformed predictions (orange), and the predictions with additional scatter (yellow) for the \HI absorber overdensities and temperatures. The predicted physical conditions from the RF models are clearly too closely concentrated towards the mean. In contrast, applying the normalising transform approach results in a predicted dataset which closely matches the truth data.
When additional scatter is instead added to the predictions, the resulting distribution also more closely matches the truth data, although not so precisely. Quantitatively, a two-sample Kolmogorov–Smirnov test with respect to the truth data gives $p$-values of $>0.95$ for the transformed overdensity and temperature distributions, but $\sim 0.25$ for the distributions with added scatter.

There are pros and cons to including the normalising transform approach to ensure that the phase space scatter is well reproduced. If one wanted to compute distribution functions for physical quantities inferred from absorption line data, not including this post-processing step would result in the distribution functions improperly capturing the tails, which may be important for some applications. \response{However, the additional step necessarily degrades the $\sigma_{\perp,\ {\rm norm}}$ and MSE$_{\rm\ norm}$ of the predictions, albeit only marginally. The correlation coefficient does not change since we are only scaling the predictions. For example, in the case of \HI\ overdensity, the $\sigma_{\perp,\ {\rm norm}}$ and MSE$_{\rm\ norm}$ increase from 0.38 $\rightarrow$ 0.40 and 0.22 $\rightarrow$ 0.25 respectively after applying the normalising transform. For \HI temperature, the $\sigma_{\perp,\ {\rm norm}}$ and MSE$_{\rm\ norm}$ increase from 0.33 $\rightarrow$ 0.34 and 0.13 $\rightarrow$ 0.14, respectively.}
Whether or not to employ the above method thus depends on the application.

\response{To recap, the normalizing transform approach requires setting up an inverse normalizing transform from the training data.  To apply this in practice, one then runs the ML pipeline on the input features, produces a predicted distribution, normalizes this distribution using the quantile transformer, and then applies the inverse transform from the training data.  This then gives predicted values that includes the additional scatter required to reproduce the spread in the input data.  Indeed, the normalizing transform steps can be thought of as part of the ML pipeline itself in order to reproduce closest to the true distribution as possible.}

\section{Conclusions}\label{sec:conclusions}

We have produced machine-learnt mappings between CGM absorption observables and the underlying gas conditions for \HI\ and selected metal lines using a Random Forest approach. RF models are preferred over other ML techniques for their relative simplicity and interpretability. These mappings represent a proof of concept for using ML models as part of an analysis pipeline for observational CGM data, which crucially does not make simplifying assumptions about the phase or composition of the absorbing gas. We identify a general tendency of the RF models to output a narrower predicted distribution than in the input data. We demonstrate two methods of reproducing the scatter of the input data: first by adding random Gaussian noise to the predictions, and second by transforming the predictions to the shape of the truth data.
Our main results are as follows:

\begin{itemize}

    \item The RF models predict reasonable \HI\ overdensities \response{($\sigma_{\perp,\ {\rm norm}} = 0.38$ dex, MSE$_{\rm\ norm}$ = 0.22) and temperatures ($\sigma_{\perp,\ {\rm norm}}= 0.33$ dex, MSE$_{\rm\ norm}$ = 0.13)}. The predictions of overdensity and temperature are highly correlated with their truth values.
    Metallicity is less well predicted \response{($\sigma_{\perp,\ {\rm norm}}= 0.42$ dex, MSE$_{\rm\ norm}$ = 0.42);} metallicity is not directly traced by HI, therefore the learned relationship likely arises from the correlation with density.

    \item The RF models also predict reasonable metal absorber conditions and perform to a similar accuracy among all metal lines, with median \response{$\rho_r = 0.68, 0.71, 0.68$, median $\sigma_{\perp,\ {\rm norm}} = 0.52, 0.44, 0.52$ and median MSE$_{\rm\ norm} = 0.24, 0.13, 0.18$ for the overdensity, temperature and metallicity predictions, respectively. }
    
    \item We report a bimodality in the absorber metallicity distributions for four of the five metal lines (\CII, \SiIII, \CIV\ and \OVI), suggesting multiple origins for the CGM gas in the Simba model.

    \item In terms of feature importances, the RF models learn \HI\ absorber overdensity from column density and equivalent width, temperature from the Doppler parameter, and metallicity from the LOS velocity separation. Low ion feature importances are similar to \HI, except that metallicities are learned from sSFR and $\kappa_{\rm rot}$. High ion feature importances are similar to the low ions, except that the overdensities are learned from radial distance and galaxy properties.
    
    \item In terms of predictive accuracy, the radial distance and LOS separation provide the most useful information for predicting \HI\ overdensity; the radial distance is also most useful for the metal line overdensities. The Doppler parameter is again the most important feature for predicting temperature for all lines. The LOS separation provides the most useful information for predicting \HI\ metallicity; the predictions for all lines are degraded by removing galaxy properties.

    \item The predictions for overdensity and temperature reproduce the phase space structure seen in the original data for all six ions, despite being trained for separately in the RF models. This verifies that accurate predictions can be made for multiple physical properties per observation.
    
    
    \item By mapping the predicted data distributions onto the shape of the input distributions using a quantile normalising transformer, we can reproduce the intrinsic scatter in the CGM phase space conditions with no loss of predictive accuracy or phase space structure.

\end{itemize}

Although we have considered \HI\ and the metal ions separately, future work on this topic could explore RF models using combinations of absorption lines to assess whether predictions may be improved by using information from multiple ion species.
A shortcoming of the ML models presented here is the that the predictions are too concentrated towards the mean of the distribution. Further development would be needed on the pipeline in order to reproduce the scatter in the original data without losing information in phase space. 

The motivation of this project is to develop a useful analysis tool for the astronomical community to aid in interpreting absorption observations of the CGM, assuming the galaxy formation model of the \simba\ simulations and the \cite{faucher-giguere_2020} UVB. The next phase of this work is to test the method by applying the ML mappings to real observational data and comparing to results derived from ionisation modelling. As such, the trained models produced for this work are available online and we encourage others to test the RF models on their own observational data. 

A natural extension of this project will be to develop additional ML mappings using absorber data from other simulations such as EAGLE and IllustrisTNG to assess the impact of galaxy formation models on the predicted conditions for the CGM.  Training the RF models on data from one simulation and testing on data from another would provide a robust test of the impact of galaxy formation model on our results. In addition, developing mappings using absorbers from the CAMELS project \citep{villaescusa-navarro_2021} would test the dependence of our results on both astrophysical and cosmological models.

\section*{Acknowledgements}

We acknowledge helpful discussions with Arif Babul, and thank Philip Hopkins for making \gizmo\ public, Robert Thompson for developing \caesar, and Horst Foidl, Thorsten Naab and Bernhard Roettgers for developing \pygad.
SA is supported by a Science \& Technology Facilities Council (STFC) studentship through the Scottish Data-Intensive Science Triangle (ScotDIST).
RD acknowledges support from the Wolfson Research Merit Award program of the U.K. Royal Society.
Throughout this work, DS was supported by the STFC consolidated grant no. RA5496 and by the Swiss National Science Foundation (SNSF) Professorship grant no. 202671. 
CCL acknowledges support from a Dennis Sciama fellowship funded by the University of Portsmouth for the Institute of Cosmology and Gravitation.

\simba\ was run on the DiRAC@Durham facility managed by the Institute for Computational Cosmology on behalf of the STFC DiRAC HPC Facility. The equipment was funded by BEIS (Department for Business, Energy \& Industrial Strategy) capital funding via STFC capital grants ST/P002293/1, ST/R002371/1 and ST/S002502/1, Durham University and STFC operations grant ST/R000832/1. DiRAC is part of the National e-Infrastructure. 

For the purpose of open access, the author has applied a Creative Commons Attribution (CC BY) licence to any Author Accepted Manuscript version arising from this submission.

\section*{Data Availability}

The simulation data and galaxy catalogs underlying this article are publicly available\footnote{\url{https://simba.roe.ac.uk}}. The software used in this work is available on Github\footnote{\url{https://github.com/sarahappleby/cgm_ml}} and the derived data will be shared on request to the corresponding author.



\bibliographystyle{mnras}
\bibliography{main} 



\appendix

\section{Results for other metals}\label{sec:additional_metals}

\begin{figure*}
	\includegraphics[width=\textwidth]{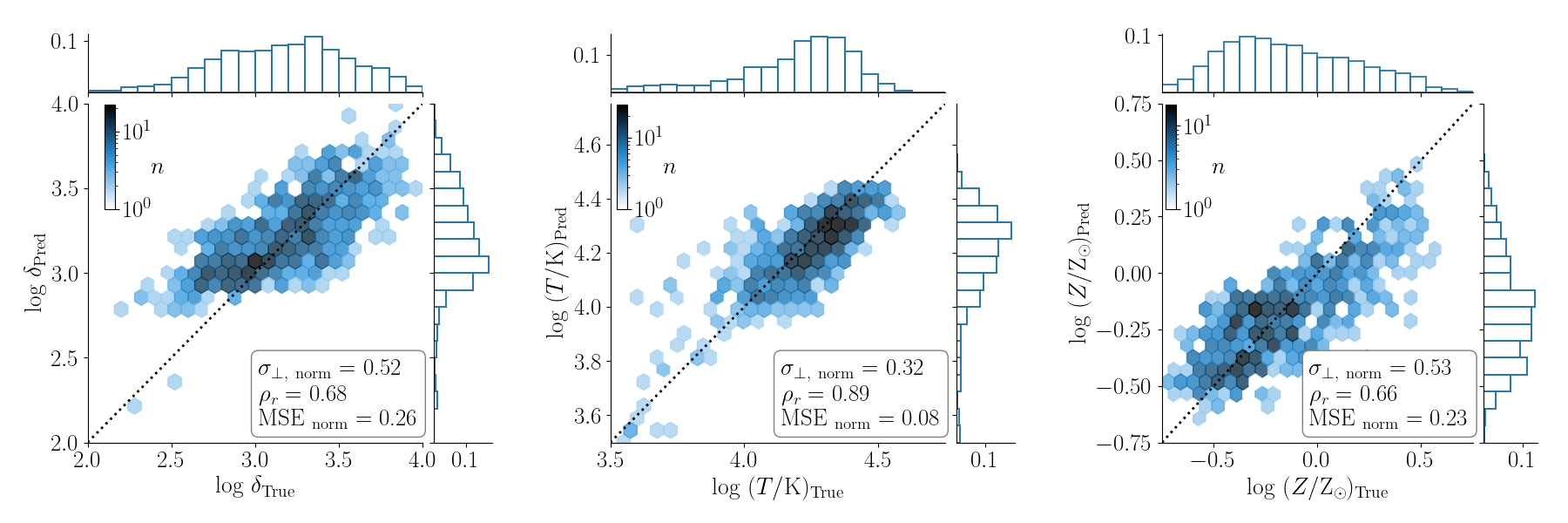}
	\vskip-0.1in
    \caption{As in Figure \ref{fig:hi_joint}, showing the predictions and true values for \MgII\ absorbers.} 
    \label{fig:mgii_joint}
\end{figure*}

\begin{figure*}
	\includegraphics[width=\textwidth]{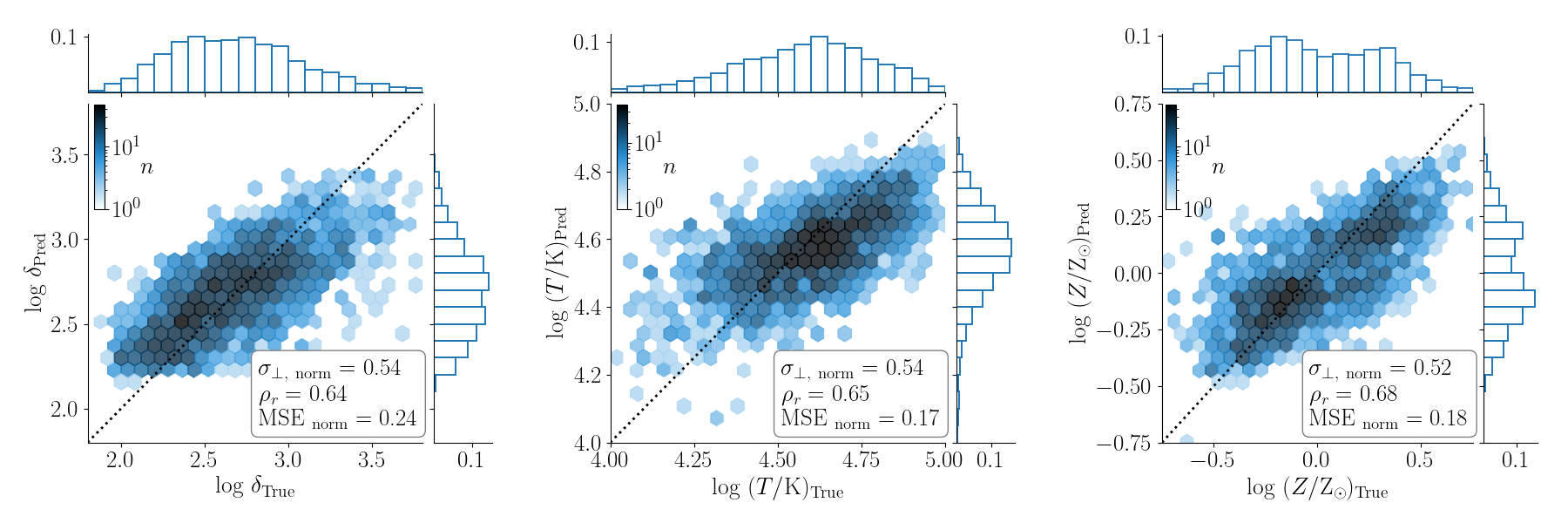}
	\vskip-0.1in
    \caption{As in Figure \ref{fig:hi_joint}, showing the predictions and true values for \SiIII\ absorbers.} 
    \label{fig:siiii_joint}
\end{figure*}

\begin{figure*}
	\includegraphics[width=\textwidth]{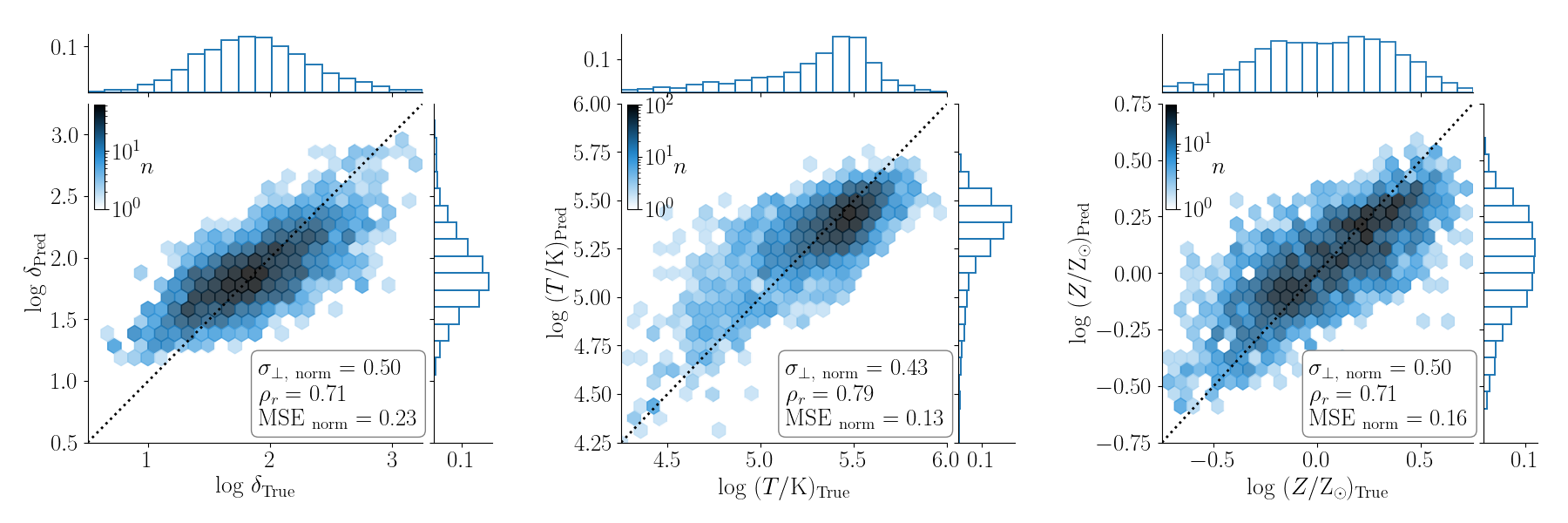}
	\vskip-0.1in
    \caption{As in Figure \ref{fig:hi_joint}, showing the predictions and true values for \OVI\ absorbers.} 
    \label{fig:ovi_joint}
\end{figure*}

For completeness, here we present figures similar to Figure~\ref{fig:hi_joint}, for \MgII, \SiIII, and \OVI.  The general trends are already captured by the plots in the main text for \CII\ and \CIV.  However, there are some interesting notable point.  For instance, \OVI\ shows significantly higher temperatures, as expected since it is a higher ionisation line.  The metallicity bimodality is still slightly present for \OVI, although at a much lower significance than for the lower ions.  \SiIII\ has the largest scatter in the recovered $T$, and also shows some bias such that high-$T$ absorbers are under-predicted while low-$T$ ones are overpredicted.  This may be because \SiIII\ seems to have absorbers spanning the widest range in temperatures from among the metal ions considered.  In terms of the RF performance, however, these ions tell a similar story, which is encouraging since it means the RF methodology is widely applicable with similar efficacy across a range of commonly observed low-$z$ UV ions.

\section{Direct Gaussian approach to adding phase space scatter}

\begin{figure*}
	\includegraphics[width=\textwidth]{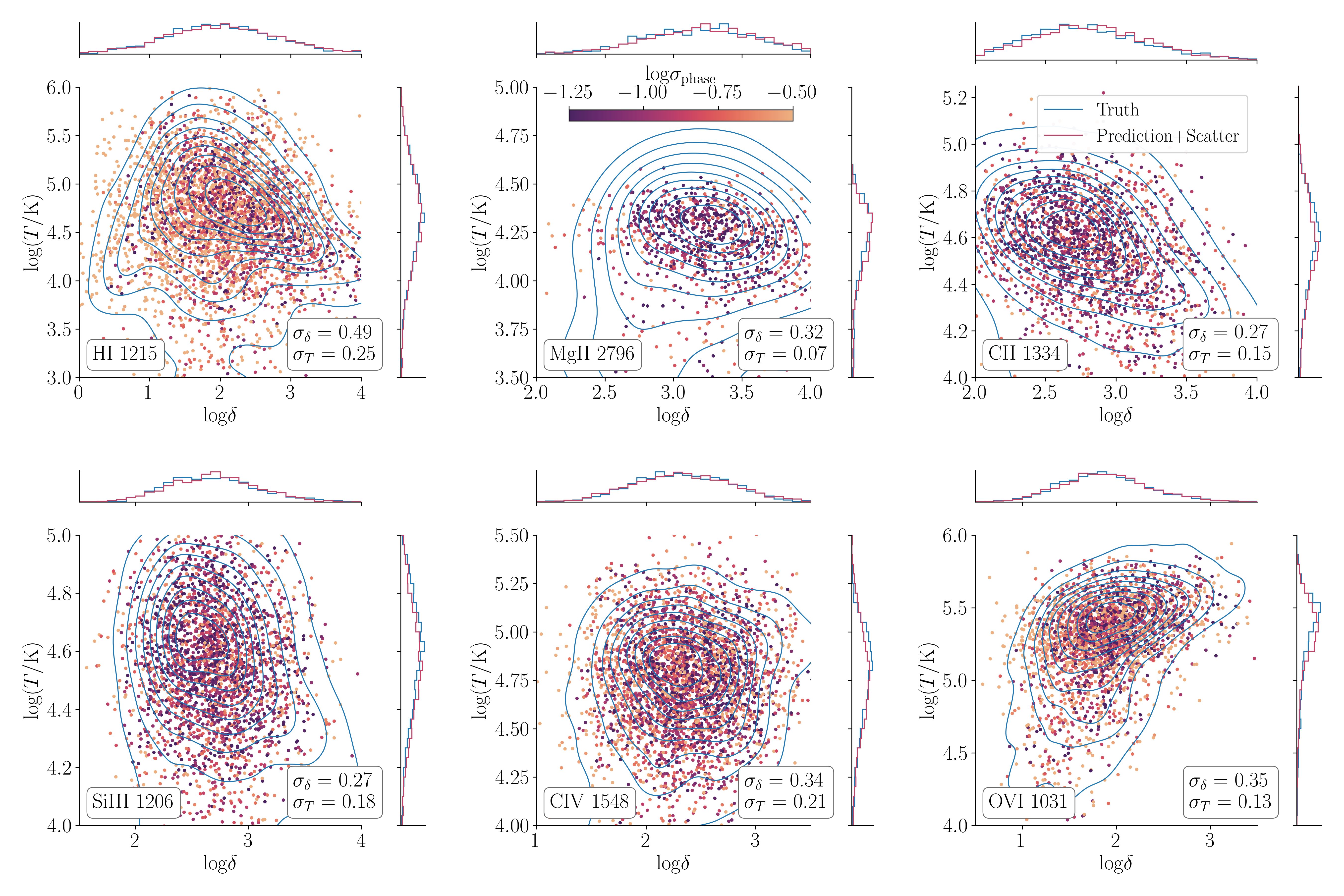}
	\vskip-0.1in
    \caption{Predicted temperature against predicted overdensity for each of the 6 ions we consider, with added random Gaussian noise to reproduce the original 1D distributions. Points are coloured by overall phase space fractional error. The 1D truth and predicted distributions are shown along the top and right of each panel. The contours show the true distribution in phase space for the test dataset. The widths of the random noise Gaussians for $\delta$ and $T$ are shown in the bottom right of each panel.  Compared to the normalising transform results shown in Figure~\ref{fig:phase_space_trans}, this approach washes out features in phase space substantially more.} 
    \label{fig:phase_space_scatter}
\end{figure*}

In \S\ref{sec:phase_space} we described our normalising transform-based approach to adding scatter to the RF predictions in order to match the 2-D truth distributions in phase space.  A more straightforward approach is to directly add Gaussian scatter to the predicted $\delta$ and $T$ distributions to match the truth without first applying a normalising transformation.  However, the results were less satisfactory.

Figure~\ref{fig:phase_space_scatter} shows the results, which can be compared to Figure~\ref{fig:phase_space_trans}.  It is clear that the simpler approach causes features within the true phase space to be more washed out, and substantially degrades the predictive accuracy.  Thus we prefer the normalising transform approach presented in \S\ref{sec:phase_space}.


\bsp	
\label{lastpage}
\end{document}